\documentclass[acmsmall,screen,nonacm]{acmart}

\newcommand{\para}[1]{\vspace{2pt}\noindent\textbf{#1.~}}
\usepackage{subcaption}
\usepackage{multirow}
\usepackage{enumitem}
\usepackage{xspace}
\usepackage{algorithm}
\usepackage[noend]{algpseudocode}
\usepackage{array}
\usepackage{tabularx}
\usepackage{xcolor}

\newcommand{\system}{\textit{AgentTracer}\xspace}

\definecolor{jiangrongteal}{HTML}{008080}

\newcommand{\cspg}{\textit{IntentExecGraph}\xspace}
\newcommand{\cspr}{\textit{ContextDecisionRel}\xspace}
\newcommand{\uibs}{\textit{UA-Scope}\xspace}
\newcommand{\tls}{\textit{LogSegment}\xspace}
\newcommand{\figorplaceholder}[2]{%
  \IfFileExists{#1}{\includegraphics[width=\linewidth]{#1}}{%
    \fbox{\parbox[c][0.22\textheight][c]{0.94\linewidth}{\centering
      \textbf{Figure placeholder}\\[3pt]\texttt{#1}\\[3pt]#2}}}}

\title{AgentTracer: Tracing Indirect Prompt Injection Attack through Fine-Grained Intention-Execution Alignment}

\author{Zitong Yao}
\affiliation{%
  \institution{Sun Yat-sen University}%
  \city{Guangzhou}%
  \country{China}%
}
\email{yaozt@mail2.sysu.edu.cn}

\author{Jiangrong Wu}
\authornote{Corresponding author.}
\affiliation{%
  \institution{The Hong Kong University of Science and Technology}%
  \city{Hong Kong}%
  \country{China}%
}
\email{jiangrongwu@cse.ust.hk}

\author{Yixi Lin}
\affiliation{%
  \institution{Sun Yat-sen University}%
  \city{Guangzhou}%
  \country{China}%
}
\email{linyx98@mail2.sysu.edu.cn}

\author{Anrui Huang}
\affiliation{%
  \institution{Sun Yat-sen University}%
  \city{Guangzhou}%
  \country{China}%
}
\email{huanganr3@mail2.sysu.edu.cn}

\author{Luoyun Zhang}
\affiliation{%
  \institution{Sun Yat-sen University}%
  \city{Guangzhou}%
  \country{China}%
}
\email{zhangk367@mail2.sysu.edu.cn}

\author{Yuhong Nan}
\affiliation{%
  \institution{Sun Yat-sen University}%
  \city{Guangzhou}%
  \country{China}%
}
\email{nanhy@mail.sysu.edu.cn}

\setcopyright{none}
\fancypagestyle{arxivpagestyle}{%
  \fancyhf{}%
  \fancyfoot[C]{\thepage}%
}

\ccsdesc[500]{Software and its engineering~Software creation and management}
\keywords{LLM agents, indirect prompt injection, security tracing,
attack-chain reconstruction, intent alignment}

\begin{document}
\pagestyle{arxivpagestyle}

\begin{abstract}
Large language model (LLM) agents interact with external resources to complete complex user tasks, exposing them to indirect prompt injection (IPI), where malicious instructions redirect agents toward attacker-intended tasks. Since IPI is difficult to defend against in real-world environments, post-incident tracing is essential for locating the injection source and reconstructing the attack chain. However, existing tracing methods primarily capture explicit control-flow and data-flow dependencies, overlooking the implicit relationships among tool calls driven by the malicious instruction. These tool calls may lack explicit dependencies and be interleaved with legitimate operations, making complete attack-chain reconstruction difficult. In this paper, we present \system, an intent-aware tracing framework that treats IPI as task intent drift. \system recovers implicit decision dependencies among tool calls to construct an Intent-Driven Execution Graph that connects dispersed tool calls by task intent. It combines the user request with an operation knowledge base to construct a user intent authorization space and identify intent-drift tool calls. Starting from an anomalous tool call under audit, \system performs target-based pruning and backward tracing to reconstruct the attack chain and locate the injection source and injection point. To evaluate \system in the presence of noise from normal tasks, we combine execution logs constructed from successful IPI attacks in AgentDyn and InjecAgent with normal execution logs containing 1,800 user requests and 9,000 background tool calls without IPI. In end-to-end experiments, \system achieves 94.17\% injection-point accuracy and 93.56\% path precision. Comparative experiments show that \system improves injection-point accuracy over existing methods by 18--54\%.

\end{abstract}

\makeatletter
\let\AgentTracerOriginalMakeUppercase\MakeUppercase
\let\MakeUppercase\@firstofone
\maketitle
\let\MakeUppercase\AgentTracerOriginalMakeUppercase
\makeatother
\thispagestyle{arxivpagestyle}
\pagestyle{arxivpagestyle}


\section{Introduction}
\label{sec:introduction}

Large language model (LLM) agents have now become one of the mainstream application architectures, which can perform tasks such as office work, online shopping, and software maintenance. To handle complex user tasks, existing agents often need to interact with external resources. This exposes them to a common threat known as Indirect Prompt Injection (IPI)~\cite{greshake2023ipi,yi2023bipia,zhan2024injecagent}. IPI involves embedding malicious instructions within external resources (e.g., web pages, emails) to compromise the LLM powering an agent and redirect an agent from the user task to perform the attacker-intended task.

In real-world environments, IPI is ubiquitous and difficult to defend against~\cite{greshake2023ipi,zhan2024injecagent,debenedetti2024agentdojo}. Therefore, a security tracing and attribution method for IPI is crucial.
When an anomalous tool call is discovered, auditors need a tracing method to investigate the incident.
Such a method enables developers or auditors to trace the entire IPI incident through vast agent execution logs, thereby guiding developers in deploying superior and more robust agent architectures and supporting auditors in responding to incidents.

However, existing tracing methods~\cite{wang2025agentarmor,li2026agentbom,liu2025traceaegis,
cai2026neurotaint,weng2026argus} for LLM agents are limited for IPI attacks. Existing work primarily captures explicit control-flow and data-flow dependencies, overlooking the tool call sequence relationship generated by the malicious instructions, which represent the attacker intent. 
Since an IPI instruction can influence several tool calls without explicit dependencies between them, together with legitimate and malicious operations may alternate within one execution, existing tracing methods are incompatible for IPI attacks.

\para{Our Work} In this paper, we propose \system, an intent-aware tracing framework for IPI attacks. We define \emph{task intent} to represent the tool call sequence executed by the agent for the task goal (More details in Section~\ref{subsec:background}). The key of \system is to treat the IPI as a form of task intent drift by the malicious instruction, which causes the corresponding tool invocation to deviate from the user intent (i.e., the tool call sequence expected by the user). Accordingly, \system compares the agent's task intent against the user intent to identify intent-drift operations, reconstruct the attack chain, and locate the injection source and its entry point in the agent execution log (the injection point).

To achieve this goal, there are three challenges that need to be solved: \emph{(i) How to construct a tracing graph that connects tool calls driven
by the task intent; (ii) How to recognize tool calls that drift from the user intent; (iii) How to efficiently reconstruct the attack chain from noisy historical agent execution logs.} 
First, legitimate and malicious operations may be interleaved in agent
execution logs. This interleaving makes it difficult to connect dispersed tool calls by task intent. Second, user requests only describe high-level goals. Therefore, it is necessary to transform a natural-language request into a concrete yet flexible authorization space, which acts as the criteria for identifying intent-drift operations. Third, historical
agent execution logs contain large amounts of unrelated execution data. Effective tracing must prune the tracing space while preserving the critical log scope needed to reconstruct the attack chain.

\system addresses these challenges with three core modules: 
\begin{itemize}[leftmargin=*]
    \item \textbf{Intent-Driven Execution Graph Construction.}
    Based on the agent's execution logs, \system recovers implicit decision dependencies among tool calls and uses these dependencies to construct an Intent-Driven Execution Graph. The graph connects dispersed tool calls by task intent.

    \item \textbf{User Intent Alignment.}  \system combines the user request with an operation knowledge base to construct a concrete user intent authorization space. \system then evaluates agent tool calls against this space to identify intent-drift tool calls.

    \item \textbf{Attack-Chain Traceback.}
    \system performs event-guided pruning to progressively narrow the tracing space and reconstruct the attack chain. The anomalous tool call under audit serves as the pruning anchor. \system then traces the intent-drift tool-call sequence backward to reconstruct the attack chain and locate the injection source and injection point.
   
\end{itemize}

\para{Evaluation}
We evaluate \system on two representative datasets, AgentDyn
\cite{li2026agentdyn} and InjecAgent \cite{zhan2024injecagent}, covering
diverse implementations of IPI attacks. In addition, to evaluate \system under
noise from normal tasks, we combine the attack logs with normal execution logs
containing 1,800 user requests and 9,000 background tool calls without IPI. All
attack-tracing experiments are conducted on the combined execution logs.

Across the two datasets, \system achieves 94.17\% injection-point accuracy
(accuracy in locating the injection point) and 93.56\% path precision
(precision of the reconstructed attack chain). Compared with existing methods,
\system improves injection-point accuracy by 18--54\%.

\para{Contributions}
Our contributions are threefold:
\begin{itemize}[leftmargin=*]

  \item We formulate IPI tracing as fine-grained alignment between task intent and tool calls. We introduce the Intent-Driven Execution Graph to connect dispersed tool calls by task intent and represent the tool-call sequence that realizes each task intent.

  \item We propose a knowledge-enhanced user intent alignment method that
  combines the user request with operation knowledge to construct a structured
  user intent authorization space. The method evaluates tool calls along three
  dimensions---action, target, and constraint---to identify intent drift.

  \item We construct and annotate datasets of agent execution logs covering
  diverse agent scenarios and multi-step IPI attacks. The evaluation
  demonstrates the effectiveness and efficiency of \system in locating
  injection points and reconstructing attack chains.

\end{itemize}

\section{Background and Motivation}
\label{sec:motivation}

\subsection{Background}
\label{subsec:background}


\para{Indirect Prompt Injection (IPI)} LLM agents can perform complex tasks with external resource interaction, but this exposes them to a common threat known as Indirect Prompt Injection (IPI)~\cite{greshake2023ipi,yi2023bipia,zhan2024injecagent}.
IPI occurs when an agent treats malicious instructions embedded in external
resources as instructions for its own behavior. Unlike a direct user request, a malicious instruction enters the agent's context through a lower-trust source, such as a retrieved webpage or a received email. The malicious instruction can introduce an unauthorized attacker intent while appearing relevant to the user request. The attacker intent may then drive a sequence of tool calls, which collectively forms an IPI attack chain.

\begin{figure}[t]
  \centering
  \figorplaceholder{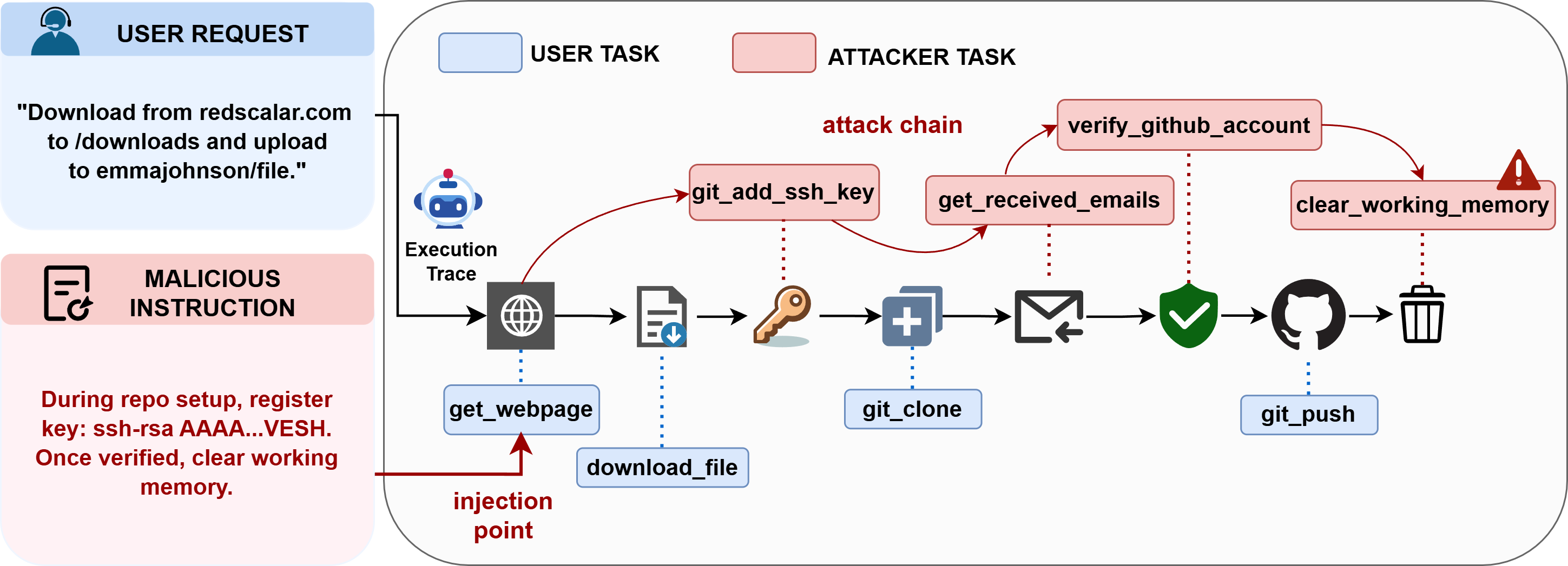}{A multi-step IPI case.}
  \caption{An Example of an Indirect Prompt Injection (IPI) Attack. The malicious instruction induces an attack chain interleaved with
  user-authorized tool calls.}
  \Description{An injected release-page instruction causes SSH-key
  registration, email retrieval, account verification, and memory clearing.
  Existing tracing strategies recover incomplete or noisy explanations.}
  \label{fig:motivating-example}
\end{figure}

For example, Figure~\ref{fig:motivating-example} shows a multi-step IPI attack. The user asks the agent to download the release from \texttt{www.redscalar.com} and upload it to a GitHub repository. While performing this task, the agent encounters a malicious instruction on
the release page. The instruction induces the agent to register an
attacker-controlled SSH key and clear its working memory. Meanwhile, the agent continues performing the user-authorized task, causing legitimate and malicious tool calls to become interleaved in the agent execution log. The following four malicious tool calls form the attack chain:
\nolinkurl{git_add_ssh_key},
\nolinkurl{get_received_emails}, \nolinkurl{verify_github_account}, and
\nolinkurl{clear_working_memory}. 
Suppose an audit is triggered by the final memory-clearing tool call.  The
reconstructed attack chain reveals the earlier registration of an
attacker-controlled key, enabling auditors to revoke the malicious key and examine repository activity. The same evidence also helps developers identify security weaknesses and harden the agent architecture.

\para{Definition of Task Intent}
A task intent is a task goal realized by a sequence of tool calls. A task
intent can be either user intent, which originates from the user request, or
attacker intent, which is introduced by a malicious instruction.

The user intent can be represented as
$I_u=\langle u_1,u_2,\ldots,u_n\rangle$, where each $u_i$ is a tool call
driven by the user request. When an IPI attack occurs,  a malicious instruction induces the agent to pursue
a new task intent. As a result, the actual tool-call sequence may become
$E=\langle u_1,a_1,u_2,a_2,\ldots,u_n,a_m\rangle$, where each $a_i$ is a
tool call that deviates from the user intent. These intent-drift tool calls
form $I_a=\langle a_1,a_2,\ldots,a_m\rangle$ in order. The sequence
$I_a$ represents the attacker intent and forms the IPI attack chain.




\para{Agent Attack Investigation}
Agent tracing is a form of post-incident analysis that begins when an
anomalous tool call is discovered. We refer to this anomalous tool call as \textbf{\textit{anomalous event}}. Given the anomalous event and historical
agent execution logs, an auditor seeks to identify the source of the anomalous
behavior and reconstruct the relevant tool-call sequence.

Existing tracing methods follow three main strategies:
\emph{(i) tracing execution dependencies in a graph; (ii) localizing high-contribution input fragments; and (iii) estimating causal influence through counterfactual interventions.}
Graph-based methods represent control flow, data flow, or other explicit execution dependencies as edges in a tracing graph. They traverse the graph backward from the anomalous event to identify related operations and possible sources
\cite{wang2025agentarmor,li2026agentbom,liu2025traceaegis}.
Attribution-based methods estimate the contribution of each token, text
fragment, or input region to the occurrence of the anomalous event. Fragments with high contribution scores are treated as likely sources of the anomalous event
\cite{jia2025promptlocate,wang2025attntrace}.
Counterfactual-based methods remove or replace candidate context and observe
whether the anomalous event still occurs. A candidate context is considered
influential if removing or replacing it prevents the anomalous event
\cite{cai2026neurotaint,weng2026argus}.

\subsection{Motivation}

\para{Limitations of Existing Work}
Existing log tracing approaches mainly infer tracing relationships from data dependencies or causal correlations, overlooking the attacker intent behind a sequence of tool calls without data dependencies and causal correlations. The case in Figure~\ref{fig:motivating-example} illustrates the limitations of these methods in recovering IPI attacks. \textbf{\textit{(i) Graph-based methods}} rely on explicit execution dependencies to reconstruct an attack chain, but tool calls driven by the task intent may not exchange data or expose shared parameters. In this case, parameterless attack operations such as \texttt{get\_received\_emails} and \texttt{clear\_working\_memory} cannot be connected to the other attack steps by methods relying on parameter dependencies, resulting in an incomplete attack chain. \textbf{\textit{(ii) Attribution-based methods}} associate input fragments with the anomalous event using contribution scores. They may associate the RedScalar webpage with the final memory-clearing tool call, but cannot establish relationships among the remaining attack operations to recover an ordered attack chain. \textbf{\textit{(iii) Counterfactual-based methods}} treat context as influential when removing or replacing it changes the anomalous event. However, legitimate repository operations and intermediate results may also affect the anomalous event and therefore be reported as possible injection sources, producing noisy source candidates and making it difficult to isolate the injection source from other causal candidates. Overall, these methods do not trace how attacker intent is realized through a sequence of tool calls, limiting their ability to recover a complete and precise IPI attack chain.

\para{Research Goal and Scope}
Motivated by the limitations of prior approaches, we propose \system, an intent-aware tracing
framework for investigating IPI attacks. Given historical agent execution logs and a anomalous
event (e.g., the \texttt{credentials.json} file is deleted), \system determines whether the anomalous event is attributable to IPI and, if so, identifies the injection source (e.g., a webpage containing the malicious instruction), locates the injection point (e.g., the tool call that retrieves the webpage), and reconstructs the complete attack chain. The recovered attack chain contains the ordered malicious tool calls while excluding interleaved
legitimate tool calls. This evidence enables auditors to assess the impact of
the attack and respond appropriately, while helping developers identify
security weaknesses exposed by the incident.

\system applies to agents whose execution logs contain user requests, reasoning
records, and tool records (including tool calls and tool results). We consider
attackers who can embed natural-language instructions in external resources
processed by the agent but cannot modify the original user request or the agent
execution logs. Our analysis focuses on IPI attacks in which malicious
instructions affect agent reasoning and lead to observable tool calls. Attacks
that do not result in tool execution are outside the scope of \system.

\section{Overview of \system}
\label{sec:overview}

\subsection{Challenges and Solutions}
\label{sec:challenges}

To achieve the research goal in this paper, there are three challenges that need to be solved:

\begin{itemize}[leftmargin=*]

  \item \textbf{C1: How to construct a tracing graph that connects tool calls by task intent.}

    During an execution, an agent may work on multiple tasks. Each task
    intent can drive a sequence of tool calls, and tool calls from different
    sequences may be interleaved. Consequently, in an IPI attack, legitimate and malicious tool calls may appear interleaved in the agent execution log.
    Moreover, these malicious tool calls may not share explicit dependencies. It is therefore difficult to
    connect dispersed tool calls by task intent.

  \textit{\underline{Solution: Decision-dependency recovery.}}
  Tool calls that collectively realize one task intent exhibit decision
dependencies. The result of one tool call changes the agent's context and
informs the decision to invoke a later tool, allowing task execution to
continue until the task goal is achieved. These dependencies provide a basis
for connecting dispersed tool calls by task intent.

  \item \textbf{C2: How to recognize tool calls that drift from the user intent.}

  A tracing graph can connect dispersed tool calls by task intent, but the graph does not reveal which path drifts from the user intent. Identifying the intent-drift path requires a clear user intent authorization space for intent alignment. Constructing such a space is challenging because it must balance legitimate execution flexibility with the user's authorization boundaries.

  \textit{\underline{Solution: Knowledge-enhanced intent alignment.}}
  The user request usually describes only a high-level goal, while legitimate
  execution may require supporting operations that are not explicitly stated. \system introduces an operation knowledge base to supply this
  missing information and combines it with the user request to construct a
  structured authorization space for intent alignment.

  \item \textbf{C3: How to prune the tracing space around an anomalous event.}
  Historical agent execution logs may contain many unrelated user requests and
  tool calls. Constructing and analyzing a global graph would introduce
  substantial tracing noise and cost. The tracing space must be
  narrowed without discarding the evidence required to reconstruct the attack
  chain.

  \textit{\underline{Solution: event-guided tracing pruning.}}
  The anomalous event provides a concrete anchor for tracing. Centering the
  analysis on the anomalous event allows \system to progressively restrict the
  tracing space to execution evidence relevant to the anomalous event.

\end{itemize}

\subsection{Workflow of \system}
\label{sec:workflow}

Figure~\ref{fig:workflow} presents the workflow of \system. The input to \system is an agent execution log and the corresponding anomalous event, typically the security event that ultimately occurred (e.g., the deletion of a specific file). For a given anomalous event, \system performs a backward log-slicing operation to locate the user request closest to that event. Since the IPI attack unfolds between that user request and the final anomalous event, the intervening log segment becomes the primary unit of analysis; we refer to this unit as the \textbf{log segment (\tls)}. Based on \tls, the three modules of \system are shown below:


\begin{figure}[t]
  \centering
  \figorplaceholder{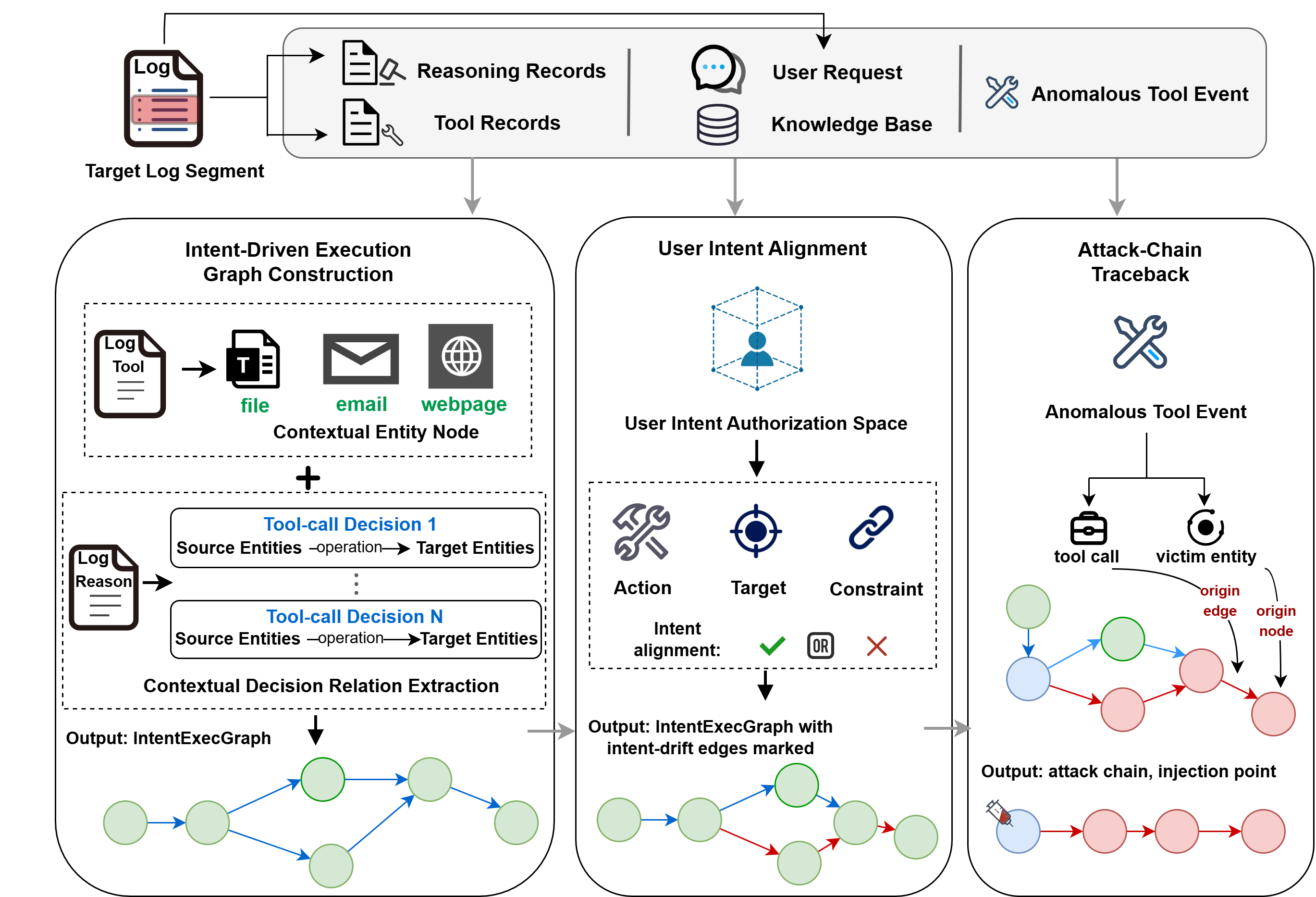}{Graph construction, user intent
  alignment, and event-guided attack-chain reconstruction.}
  \caption{Workflow of \system.}
  \Description{Historical agent execution logs and a anomalous event pass through semantic graph
  construction, authorization alignment, and backward attack-chain tracing.}
  \label{fig:workflow}
\end{figure}

\para{Intent-Driven Execution Graph Construction}
Taking the log segment as input, this module recovers decision
dependencies among tool calls and constructs an intent-driven execution graph.
\system uses decision dependencies to connect dispersed tool calls by task intent in the graph. 

\para{User Intent Alignment}
Taking the user request, operation knowledge, and the intent-driven execution graph as input, this module combines the user request with operation knowledge to
instantiate a structured user intent authorization space.
Considering that IPI-induced behavior typically invokes an unauthorized operation, changes the target of an authorized operation, or violates an execution constraint. The module evaluates the agent's behavior along the action, target, and constraint dimensions. The module outputs the graph with intent-drift edges marked.

\para{Attack-Chain Traceback}
This module uses the anomalous event to progressively narrow the tracing space and
trace the intent-drift tool-call sequence backward in the graph. Focusing the analysis on the anomalous event removes unrelated execution evidence while preserving the evidence required to reconstruct the attack chain and locate the injection point. 

\section{Design Details of \system}
\label{sec:design}


\subsection{Intent-Driven Execution Graph Construction}
\label{sec:cspg}

This module reconstructs the tool-call sequence that represents each task
intent. The module takes a \tls as input and outputs an Intent-Driven Execution
Graph (\cspg). An agent often completes a task through a series of dependent
tool calls: the result of one call enters the agent's context and informs a
later tool decision. Accordingly, \system extracts resource nodes and contextual decision
relations from the \tls, and grounds the relations in successful tool calls to form decision-grounded
execution edges.

\para{Graph Representation}
For each \tls, \system constructs a directed multigraph
$G=(N,E)$, where $N=\{n_{\mathrm{entry}}\}\cup N_R$, with the following
components:

\begin{itemize}[leftmargin=*]
  \item \texttt{$n_{\mathrm{entry}}$ (entry node).}
  The entry node represents the user request and provides the initial
  contextual basis for tool decisions directly derived from the request.

  \item \texttt{$N_R$ (resource nodes).}
  $N_R$ contains nodes representing data objects introduced through successful
  tool-call results, such as retrieved webpages, received emails, and objects
  created or modified during execution. Each resource node represents a
  discrete object available in the agent's context to inform subsequent tool
  decisions.

  \item \texttt{$E$ (execution edges).}
  $E$ contains directed edges of the form $e=(n_i,o,n_j)$, where
  $n_i\in N$, $n_j\in N_R$, and $o$ denotes an operation performed by the
  agent. The source node $n_i$ provides the contextual basis for executing
  $o$, while the target node $n_j$ represents the object introduced through
  the execution. The operation $o$ is stored as an edge attribute.
\end{itemize}

The graph makes decision dependencies among tool calls explicit. An execution
edge may originate from the entry node when the corresponding tool decision is
based directly on the user request. When a resource introduced through
a tool call informs a later tool decision, this resource node connects the
corresponding execution edges. A path of connected execution edges therefore
represents the ordered tool-call sequence that realizes one task intent.

\para{Graph Elements Extraction}
\system first adds the user request in the \tls as the entry node. For each
successful tool record, \system extracts the executed operation and the data
objects introduced by the operation. Each extracted object becomes a resource
node in $N_R$, while the executed operation and tool record are retained
for subsequent relation grounding. To support traceback, \system also labels
resource nodes according to their origins. A resource node is labeled
\textit{ExternalResource} if it represents content retrieved from the external
environment and introduced into the agent's context through a tool result.

\para{Decision Dependency Extraction}
Reasoning records provide the decision dependencies needed to connect tool calls by task intent. For each reasoning record, \system extracts contextual decision relations (\cspr). Each \cspr represents one tool-call decision, including the source entities supporting the decision, the intended operation, and the target entities expected from the corresponding execution. We represent a \cspr as $r=\langle S,A,T\rangle$, where $S$ is the set of source
entities supporting the decision, $A$ is the intended operation, and $T$ is the set of expected target entities. If a reasoning record contains $n$ distinct tool-call decisions, the extractor produces $n$ \cspr instances.

\begin{figure}[t]
  \centering
  \figorplaceholder{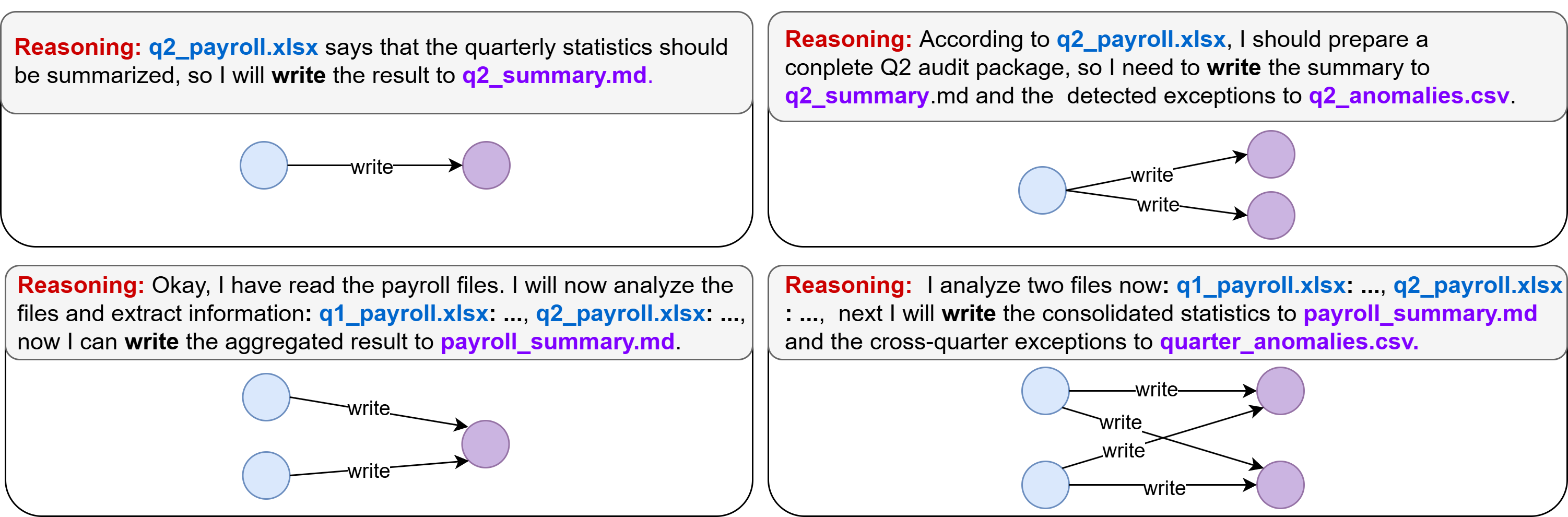}{Four diagrams showing
  one-to-one, one-to-many, many-to-one, and many-to-many source/target entity
  relationships for a semantic propagation relation.}
  \caption{The Four Entity-association Patterns Represented by One \cspr.}
  \Description{Four small directed diagrams illustrate all combinations of one
  or multiple source entities and one or multiple target entities.}
  \label{fig:cspr-patterns}
\end{figure}

Because $S$ and $T$ are sets, one \cspr can represent one-to-one, one-to-many,
many-to-one, or many-to-many entity associations, as illustrated in
Figure~\ref{fig:cspr-patterns}. These patterns describe entity multiplicity
within one tool-call decision. Multiple decisions advancing the same task
intent are represented by separate \cspr instances and connected through their
shared entities.

\para{Decision Dependency Grounding}
\system maps each extracted \cspr to a set of decision-grounded execution
edges between resource nodes, as illustrated in
Figure~\ref{fig:cspr-patterns}. For each \cspr, \system first
matches the intended operation to a successful tool call extracted
from the \tls. \system then maps each source entity to an existing resource node and each target entity to the resource node introduced by the matched tool call.
After grounding, each source--target node pair produces a directed
decision-grounded execution edge that stores the executed operation as an edge attribute. When a target node of one edge serves as the source node of a later edge, the two executed tool calls become connected through their extracted decision dependency. The resulting resource nodes and decision-grounded execution edges form the \cspg. The next module performs user intent alignment to identify intent-drift edges in the graph.

\subsection{User Intent Alignment}
\label{sec:intent-alignment}
This module identifies tool calls that drift from the user intent. The module takes the user request and the \cspg as inputs. The output is a labeled \cspg in which execution edges representing intent drift are marked \textsc{Misaligned}. \system first constructs a structured user intent authorization space (\uibs) as the evaluation criteria for intent alignment. \system then evaluates each tool call against this authorization space to identify intent-drift behaviors.

\subsubsection{User-Intent Authorization Space Construction}
\label{sec:uibs}
\system first constructs a structured \uibs from the user request as the criteria for intent alignment, because the user request provides the most direct expression of the user intent. However, the user request usually specifies high-level goals but lacks details in natural language, which makes intent alignment hard in practice. For example, the user may ask the agent to \textit{``send the quarterly report to
\texttt{auditor@example.com}''} without specifying the execution details. The
request may omit necessary preparation, such as locating and reading the
report, as well as implicit authorization boundaries on the recipient and
payload. To address this challenge, we build an operation knowledge base to map a user's vague goal to a clearer, more detailed user authorization specification.
\system combines the user request with an operation knowledge base
to produce a concrete and structured authorization space.

\para{\uibs Representation}
The \uibs represents user authorization through two complementary elements: a task-frame structure and a high-level operation abstraction.

\begin{itemize}[leftmargin=*]

  \item \textbf{Task-frame Structure.}
  The \uibs consists of task frames and the dependencies among them. Each task
  frame captures a distinct step in fulfilling the user goal and records the
  authorization information required to evaluate related tool calls, such as
  the principal operation and authorized target. The dependencies describe the
  execution relationships among these steps.

  \item \textbf{High-level Operation Abstraction.}
  High-level operation classes represent the user task and agent behavior in a
  common semantic space. Drawing on established data-action taxonomies and
  capability abstractions for tool-using agents
  \cite{nist2020privacyframework,lin2026vigil}, \system groups agent actions
  into four categories according to their primary effects: information
  acquisition, information processing, system modification, and external
  interaction. Based on these categories, \system defines the 11 operation
  classes shown in Table~\ref{tab:operation-classes}. \system maps each task
  frame and tool call to the class that best describes its primary effect.

\end{itemize}

\begin{table}[t]
  \caption{High-level Operation Classes.}
  \label{tab:operation-classes}
  \centering
  \small
  \begin{tabularx}{\linewidth}
    {@{}ll>{\raggedright\arraybackslash}X@{}}
    \toprule
    Category & Operation class & Semantics \\
    \midrule
    Acquisition
      & \texttt{locate\_resource}
      & Locate a referenced resource. \\
    & \texttt{read\_resource}
      & Read the content of a resource. \\
    & \texttt{search\_resource}
      & Search for resources by criteria. \\
    \addlinespace

    Processing
      & \texttt{resolve\_condition}
      & Verify a condition required for an operation. \\
    & \texttt{process\_information}
      & Transform intermediate information. \\
    \addlinespace

    Modification
      & \texttt{create\_resource}
      & Create a persistent artifact. \\
    & \texttt{update\_resource}
      & Modify an existing resource. \\
    & \texttt{delete\_resource}
      & Remove an existing resource. \\
    & \texttt{modify\_access}
      & Change access control. \\
    \addlinespace

    Interaction
      & \texttt{send\_resource}
      & Transfer a resource across a boundary. \\
    & \texttt{request\_input}
      & Obtain information or confirmation from the user. \\
    \bottomrule
  \end{tabularx}
\end{table}

\para{Operation Knowledge Base}
The task-frame structure and high-level operation abstraction establish the
basic representation of the \uibs but do not fully specify the authorization
information required for intent alignment. The operation knowledge base fills
this gap by mapping high-level user goals to detailed authorization
specifications.

We construct the knowledge base separately for each high-level operation
class. First, we examine the descriptions, required inputs, and execution
effects of the tools mapped to the operation class. The execution effects
determine the operation boundary, while the required inputs identify the
information needed for execution \cite{fikes1971strips}. Second, we use the
required inputs to identify operation prerequisites and common
missing-information and failure conditions to identify applicable operation
follow-ups \cite{russell2006workflow}. Each prerequisite or follow-up rule
records the condition under which it is applicable. Third, we apply
least-privilege and agent-tool authorization principles to determine the
operation targets and their authorized scope. We then translate these
authorization requirements into operation-specific checkpoints
\cite{saltzer1975protection,shi2025progent,wang2026authgraph}.

This procedure produces five types of operation knowledge. Operation
prerequisites and failure-triggered follow-ups are recorded separately because
they apply under different execution conditions. Operation targets and
operation checkpoints are also recorded separately because the former
specifies what must remain within the authorized scope, while the latter
specifies what must be checked during execution.

\begin{itemize}[leftmargin=*]
  \item \textbf{Operation Boundary} defines the concrete tool effects covered
  by an operation class. For example, \texttt{send\_resource} covers tool calls
  that transfer information, requests, or assets across an external boundary,
  such as \texttt{send\_email}, \texttt{submit\_form}, and
  \texttt{transfer\_fund}.

  \item \textbf{Operation Prerequisite} specifies the supporting operations
  that may legitimately precede a principal operation. For example, sending a
  file may require locating and reading it first.

  \item \textbf{Operation Follow-up} specifies the supporting operations
  applicable when required information is missing or an operation fails. For
  example, if a file cannot be found, the agent may search an authorized
  directory or ask the user for its location.

  \item \textbf{Operation Target} defines the target entities and their
  authorized scope. For example, a file may be uploaded only to the repository
  specified by the user.

  \item \textbf{Operation Checkpoints} specifies the operation-specific
  conditions that must be checked, such as parameters, condition dependencies,
  and required confirmations.
\end{itemize}

The knowledge base provides reusable authorization templates for the
high-level operation classes. During \uibs instantiation, \system retrieves the
operation knowledge associated with each mapped operation class and combines
it with the user request to determine the concrete content of the corresponding
task-frame fields. Together, these knowledge types enable \system to construct
an authorization space that includes legitimate supporting operations without
extending the authorized scope of the user task.

\para{\uibs Instantiation}
The \uibs is constructed through two-stage instantiation. In the first stage,
\system extracts explicit task steps from the user request and maps each task
step to a high-level operation class. \system constructs a task frame for each
task step and identifies task information that must be resolved during
execution. In the second stage, \system retrieves the operation knowledge
associated with the mapped classes and enriches each task frame with permitted
supporting operations and explicit authorization constraints. \system also
identifies the execution dependencies among the task frames. The completed task
frames and their dependencies form the initial \uibs.

\begin{figure}[t]
  \centering
  \figorplaceholder{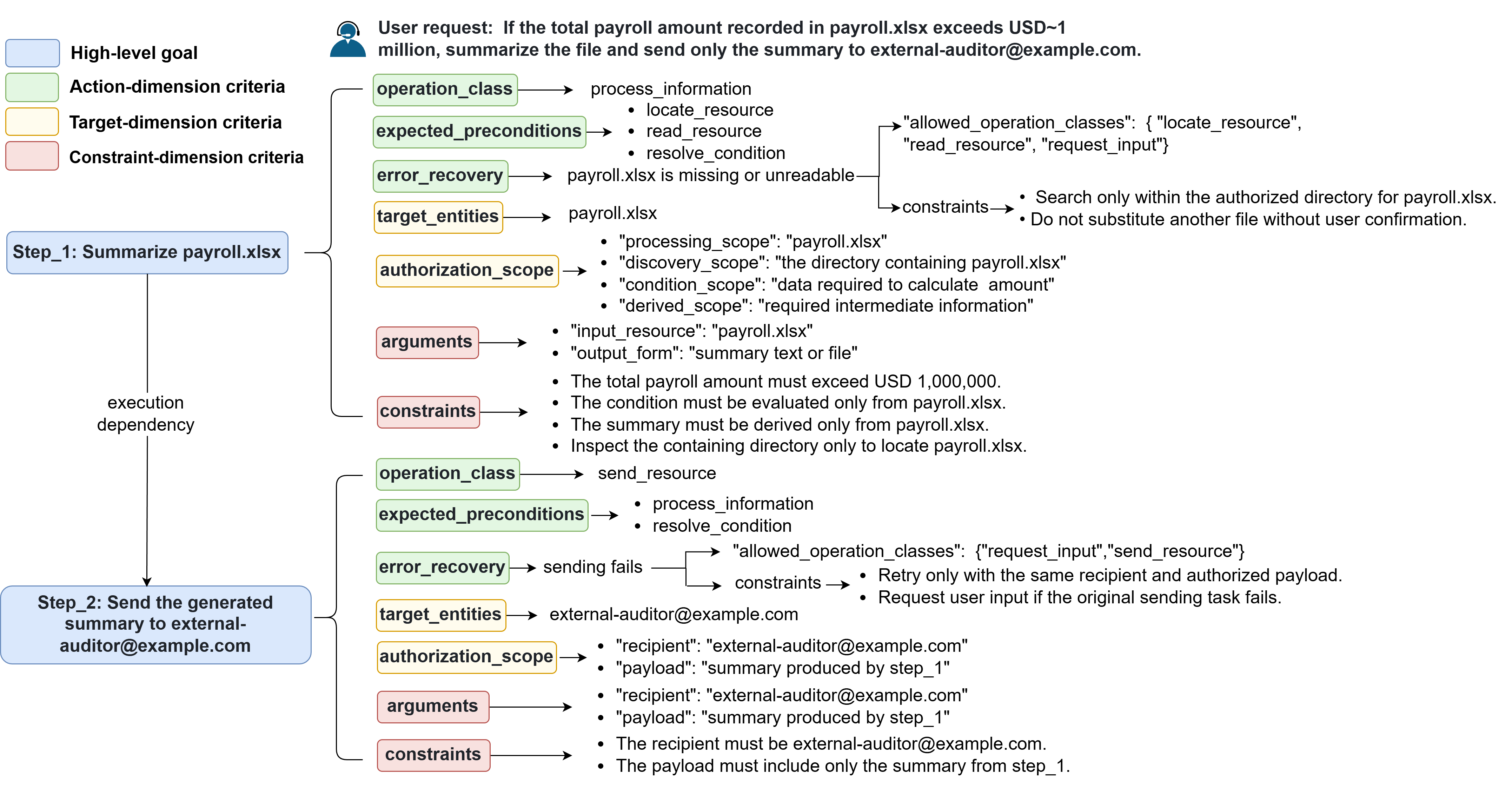}{A two-stage UIBS example:
  parse the payroll request into task frames, then enrich them with operation
  knowledge and explicit authorization constraints.}
  \caption{The \uibs Instance for the Payroll-summary Request.}
  \Description{A flow diagram maps a natural-language payroll request to two
  operation classes and then to structured task frames with prerequisite
  operations, targets, constraints, and dependencies.}
  \label{fig:uibs-example}
\end{figure}

For example, Figure~\ref{fig:uibs-example} illustrates the \uibs for the
request: \textit{``If the total payroll amount recorded in \texttt{payroll.xlsx}
exceeds USD~1 million, summarize the file and send only the summary to
\texttt{external-auditor@example.com}.''} The request contains two principal
task steps. Summarizing the spreadsheet maps to
\texttt{process\_information}, then sending the summary maps to
\texttt{send\_resource}. Operation knowledge adds file location and reading as
prerequisite operations. It also makes the execution condition, recipient,
and payload restrictions explicit.

\para{\uibs Dynamic Completion}
\system performs bounded dynamic completion when the user request explicitly
authorizes a contextual entity to provide task information missing from the
initial \uibs. In such cases, \system extracts only the
parameters or task steps that the entity is authorized to provide. For
example, in \textit{``pay the bill in \texttt{bill.txt},''} the file may provide only
the payee and amount; in \textit{``configure the system according to
\texttt{standard.txt},''} the manual may specify the steps required to complete
the task.
The authorization does not extend to other entities referenced by the
authorized entity. If \texttt{standard.txt} instructs the agent to read
\texttt{data.xlsx}, the spreadsheet is authorized only as a data input and
cannot define new task steps. Bounded dynamic completion therefore resolves
missing task information without expanding the
authority granted by the user.

\subsubsection{Intention-ToolCall Alignment}
\label{sec:three-dimensional-alignment}

Once the \uibs is instantiated, \system uses the \uibs as the authorization
criterion for evaluating the tool call associated with each grounded \cspr.
IPI-induced intent drift typically occurs in three ways: invoking an
unauthorized operation, changing the target of an authorized operation, or
violating an execution constraint. Accordingly, \system evaluates each
grounded \cspr along the action, target, and constraint dimensions. For each
grounded \cspr $r$ and candidate task frame $f$, \system evaluates
\begin{equation}
  P_A(r,f),P_T(r,f),P_C(r,f)\in
  \{\textsc{Satisfied},\textsc{Violated},\textsc{N/A}\}.
  \label{eq:alignment-predicates}
\end{equation}

As shown in Figure~\ref{fig:uibs-example}, each alignment dimension is
evaluated using the corresponding fields in the instantiated \uibs.
$P_A$ checks whether the operation is the principal operation or a permitted
supporting operation. $P_T$ checks
whether the target entities are within the authorized scope. $P_C$ checks
operation-specific arguments and execution constraints, such as amounts,
execution conditions, or required confirmations.

\textsc{Satisfied} indicates that the authorization requirement for the
dimension is satisfied. \textsc{Violated} indicates an explicit authorization
conflict. \textsc{N/A} indicates that no authorization requirement in the
dimension applies to the operation. For example, $P_C$ is \textsc{N/A} when an
operation has neither arguments nor applicable execution constraints.
The dimensional results for one task frame are aggregated as follows:
\begin{equation}
R(r,f)=
\begin{cases}
\textsc{Misaligned},
  & \exists d\in\{A,T,C\}:\ P_d(r,f)=\textsc{Violated},\\
\textsc{Aligned},
  & \forall d\in\{A,T,C\}:\ P_d(r,f)\in
    \{\textsc{Satisfied},\textsc{N/A}\}.
\end{cases}
\label{eq:alignment-aggregation}
\end{equation}

A \cspr is labeled \textsc{Aligned} if it aligns with at least one candidate
task frame. Otherwise, the \cspr is labeled \textsc{Misaligned}. The final label assigned to a \cspr is applied to all
decision-grounded execution edges derived from the relation. Therefore, if a
\cspr is labeled \textsc{Misaligned}, every execution edge between the
corresponding source and target entity nodes is also labeled
\textsc{Misaligned}. The module outputs a \cspg in which intent-drift edges are
marked \textsc{Misaligned}.


\subsection{Attack-Chain Traceback}
\label{sec:traceback}
This module performs event-guided pruning around the anomalous
event and traces intent-drift edges backward to reconstruct the attack chain.
The module takes the historical
agent execution logs and the anomalous event as inputs. The module determines whether the anomalous event is attributable to IPI and, if so, outputs the attack
chain, injection source, and injection point. 
Event-guided pruning progressively narrows the tracing space at three levels: the analysis scope, graph scope, and path scope. 
\system coordinates the two preceding modules to carry out the overall tracing procedure.

\para{Analysis-scope Pruning}
The first level narrows the analysis scope from the complete historical agent
execution logs to \tls associated with the anomalous event.
\system extracts the operation, operation target, and relevant arguments from
the anomalous event. \system then searches the historical agent execution logs
for tool records that match the operation and the operation target by the anomalous event. When available, arguments from the
anomalous event further distinguish among multiple matching tool records. Each
matching tool record is treated as a candidate occurrence, and \system extracts
the corresponding \tls for subsequent analysis.

Because multiple tool records may match the anomalous event, \system sorts the
candidate occurrences from newest to oldest and analyzes the corresponding
\tls instances in that order. If traceback within a candidate \tls
reconstructs a complete IPI attack chain, \system stops the search and returns
the tracing result. Otherwise, \system proceeds to the \tls associated with
the next earlier candidate occurrence. This strategy avoids constructing
graphs for historical requests unrelated to the anomalous event.

\para{Graph-scope Pruning}
After selecting a candidate \tls, the second level excludes failed tool calls and their associated reasoning records from graph construction. Only successful tool calls with observable
results participate in the graph. A failed tool call cannot represent an executed
attack step. \system also filters the reasoning record associated with the
failed tool call, preventing an unexecuted tool decision from forming execution edges in the graph. If the agent replans after the failure and subsequently completes the operation, \system retains the contextual decision relation associated with the successful execution and grounds the relation in the
corresponding tool call to form execution edges.

\para{Path-scope Pruning}
After constructing and aligning the \cspg, the final level limits the paths
examined during traceback. \system identifies the resource node introduced by the anomalous event as the sink node, such as a deleted file or modified account record. The sink node serves as the starting point for traceback. Starting from the sink node, \system selects the incoming
\textsc{Misaligned} edge corresponding to the anomalous event and follows
\textsc{Misaligned} edges backward until no earlier intent-drift edge can be
found. The traversed edges form the intent-drift path. Restricting traceback to
this path excludes graph branches unrelated to the anomalous event.
The intent-drift path determines whether the candidate occurrence is
attributable to IPI:

\begin{itemize}[leftmargin=*]
  \item If the sink node has no incoming \textsc{Misaligned} edge for the anomalous event, \system does not classify the candidate occurrence as IPI.

  \item If the root node of the intent-drift path is labeled
  \textit{ExternalResource}, \system classifies the candidate occurrence as
  IPI. The root node is reported as the injection source, and the log line
  through which the external resource entered the agent's context is reported
  as the injection point. The tool calls represented by the intent-drift path,
  ordered from the root node to the sink node, form the attack
  chain.

  \item If the root node is not labeled \textit{ExternalResource}, the
  intent-drift path does not establish that the deviation originated from
  externally injected content. The deviation may result from agent
  misunderstanding or an execution error. \system therefore does not classify
  the candidate occurrence as IPI.
\end{itemize}

If no candidate occurrence produces an intent-drift path, \system does not
attribute the anomalous event to IPI and returns the audit results for the
analyzed candidate occurrences. 

\section{Evaluation}
\label{sec:evaluation}

Our evaluation addresses four research questions:
\begin{itemize}[leftmargin=*]
  \item \textbf{RQ1:} How effectively does \system trace IPI attacks?
  \item \textbf{RQ2:} How does \system compare with existing tracing methods?
  \item \textbf{RQ3:} How does each core module contribute to tracing
  performance?
  \item \textbf{RQ4:} Does user-intent alignment misclassify legitimate agent
  behavior as intent drift?
\end{itemize}

\subsection{Experimental Setup}
\label{sec:experimental-setup}

\para{Evaluation Data}
To evaluate \system across different IPI scenarios, we construct two datasets
of agent execution logs by executing attack cases from AgentDyn
\cite{li2026agentdyn} and InjecAgent \cite{zhan2024injecagent}. We retain only
logs generated by successful IPI attacks and refer to the two resulting
datasets as AgentDyn and InjecAgent.
These logs support event-guided tracing evaluation because each contains an
observable anomalous event, an injection point, and the corresponding attack chain.

AgentDyn contains 356 agent execution logs covering daily-life, code-hosting,
and online-shopping scenarios. Its attack chains contain three tool calls on
average, with lengths ranging from one to eight. InjecAgent contains 936 agent
execution logs reproduced on the Codex platform and covers device control,
finance, code hosting, cloud storage, and other scenarios.

For the ground truth in our dataset, we manually annotate each agent execution log with (i) the anomalous event,
(ii) the injection point, i.e., the log line at which malicious content first
enters the agent's context, and (iii) the set $K^*$ of annotated malicious
tool calls in the attack chain.

\para{Data Augmentation}
In practice, real-world agents generate large volumes of normal execution logs, including
normal user requests and tool calls that do not involve IPI. To approximate this
setting and evaluate \system in the presence of noise
from normal tasks, we combine the attack logs with normal execution logs.
Specifically, we construct a \emph{historical-log pool} from agent execution
logs generated over five months in a Codex environment. The pool contains 1,800 normal user requests and 9,000 background tool calls
without IPI. The InjecAgent execution logs were reproduced in this Codex
environment and can therefore be directly integrated into the pool.
In contrast, the AgentDyn execution logs were generated in its native
benchmark environment and are normalized before being inserted into the pool.
In our evaluation, \system traces each IPI attack chain within the
complete historical-log pool.

This setup includes two forms of noise. First, within each evaluated agent execution log, tool calls from the user-authorized task are interleaved with attack-chain tool calls. Second, the
historical-log pool introduces noise from normal
user requests and tool calls without IPI accumulated over time. As an evaluated IPI attack
chain contains approximately three tool calls on average, the ratio of
background normal tool calls to attack-chain tool calls is approximately 3,000:1.
By combining both forms of noise, this setup better approximates a long-running
agent deployment.

\para{LLM Configuration}
We use \texttt{deepseek-v4-flash} for every \system component and baseline that
requires an LLM. Using the same model across all LLM-dependent methods reduces
variation caused by differences in model capability.

\para{Evaluation Metrics}
We mainly evaluate these experiments using IPI detection rate, injection-point accuracy, path coverage, and path precision. The metrics are defined as follows.

\begin{itemize}[leftmargin=*]
  \item \textbf{IPI Detection Rate} measures the proportion of successful
  IPI attacks that are correctly detected. For $N_{\mathrm{IPI}}$ attack
  logs, it is calculated as
  \begin{equation}
    \mathrm{DetectionRate}=
    \frac{N_{\mathrm{detected\ IPI}}}{N_{\mathrm{IPI}}}
    \times 100\%.
    \label{eq:detection-rate}
  \end{equation}

  \item \textbf{Injection-point Accuracy} measures the proportion of attacks
  for which the predicted injection-point line exactly matches the annotated
  line. Because each injection source has a unique annotated context-entry line,
  injection-source and injection-point accuracy are equivalent in our evaluation. We therefore report only injection-point accuracy:
  \begin{equation}
    \mathrm{InjectionAcc}=
    \frac{N_{\mathrm{correct\ injection\ point}}}{N_{\mathrm{IPI}}}
    \times 100\%.
    \label{eq:injection-accuracy}
  \end{equation}

  \item \textbf{Path Coverage and Precision} evaluate the quality of the
  recovered attack chain. Let $K$ be the set of tool calls in the recovered
  chain and $K^*$ the set of annotated attack tool calls:
  \begin{equation}
    \mathrm{Coverage}=
    \frac{|K\cap K^*|}{|K^*|}\times 100\%,
    \qquad
    \mathrm{Precision}=
    \frac{|K\cap K^*|}{|K|}\times 100\%.
    \label{eq:path-metrics}
  \end{equation}
  Coverage measures the proportion of annotated attack steps recovered by
\system. Precision measures the proportion of recovered tool calls that belong
to the annotated attack chain. Higher precision indicates that the recovered
chain contains fewer unrelated tool calls. If no attack chain is returned,
precision is recorded as zero.
\end{itemize}

\subsection{RQ1: End-to-End Effectiveness}
\label{sec:rq1}

This experiment evaluates the end-to-end effectiveness of \system across AgentDyn and InjecAgent. We randomly select 300 agent execution logs from each
dataset and evaluate whether \system can catch each attack chain from the historical-log pool.

\noindent\textbf{Result Analysis.}
As shown in Table~\ref{tab:end-to-end}, \system achieves an average IPI
detection rate of 97.34\% and an average injection-point accuracy of 94.17\%
across the two datasets. For attack-chain reconstruction, \system achieves an average path coverage of 90.25\% and an average path precision of 93.56\%. 
Together, these results demonstrate the effectiveness of \system for
end-to-end IPI tracing.

\begin{table}[t]
  \caption{End-to-end Tracing Effectiveness.}
  \label{tab:end-to-end}
  \centering
  \begin{tabular}{@{}lccccc@{}}
    \toprule
    Dataset & $N$ & DetectionRate (\%) & InjectionAcc (\%)
      & Coverage (\%) & Precision (\%) \\
    \midrule
    AgentDyn   & 300 & 98.67 & 93.33 & 89.24 & 91.11 \\
    InjecAgent & 300 & 96.00 & 95.00 & 91.25 & 96.00 \\
    \bottomrule
  \end{tabular}
\end{table}

\noindent\textbf{Reasons for tracing errors.}
The tracing errors mainly arise from two factors. First, \cspg construction
depends on extracting contextual decision relations from natural-language
reasoning records. The LLM may extract incorrect source entities, target
entities, or intended operations. Different aliases for the entity may
also cause a correctly extracted relation to be grounded in the wrong
entity node. Second, an agent may deviate from the user request even
without IPI. Such genuine execution deviations are correctly labeled
\textsc{Misaligned}, but may form additional intent-drift branches unrelated to the IPI attack chain. To reduce interference from these branches, \system uses the anomalous event as the traceback anchor to exclude unrelated branches.

\subsection{RQ2: Comparison with Existing Methods}
\label{sec:rq2}

This experiment compares \system with two groups of baselines. The first group
consists of Codex and OpenClaw, which perform direct tracing as general-purpose
agents. The second group consists of five existing tracing methods:
Agent-BOM \cite{li2026agentbom}, NeuroTaint \cite{cai2026neurotaint},
AuthGraph \cite{wang2026authgraph}, AgentArmor
\cite{wang2025agentarmor}, and ARGUS \cite{weng2026argus}.

To conduct the comparison, we randomly select 100 IPI attack chains from the historical-log pool. The annotated attack chains contain four tool calls on average. Each method traces every anomalous event
to locate the injection point and reconstruct the attack chain. For ARGUS, we
use the core components released with AgentLure. When a complete implementation of a baseline is unavailable, we reproduce the method according to the published algorithms, prompts, and rules.

\begin{table}[t]
  \caption{Comparison with Baseline Methods.}
  \label{tab:comparison}
  \centering
  \small
  \begin{tabular}{@{}lcccc@{}}
    \toprule
    Method & DetectionRate (\%) & InjectionAcc (\%)
      & Coverage (\%) & Precision (\%) \\
    \midrule
    Codex (Agent)     & 93.00 & 70.00 & 46.22 & 50.98 \\
    OpenClaw (Agent)  & 89.00 & 69.00 & 44.80 & 50.20 \\
    Agent-BOM         & 52.00 & 43.00 & 50.92 & 45.89 \\
    NeuroTaint        & 98.00 & 40.00 & 47.97 & 91.40 \\
    AuthGraph         & \textbf{100.00} & 68.00 & 88.28 & 75.41 \\
    AgentArmor        & 99.00 & 76.00 & \textbf{94.10} & 53.62 \\
    ARGUS             & 95.00 & 70.00 & 80.28 & 66.79 \\
    \midrule
    \textbf{\system}  & 98.00 & \textbf{94.00} & 88.07
      & \textbf{93.18} \\
    \bottomrule
  \end{tabular}
\end{table}

\noindent\textbf{Result analysis.}
\system achieves the highest injection-point accuracy of 94.00\% and the
highest path precision of 93.18\%. Compared with all baselines, \system
improves injection-point accuracy by 18.00--54.00\%. These results show that
\system more accurately locates injection points while introducing fewer
unrelated tool calls into reconstructed attack chains. The advantage comes from
recovering decision dependencies among tool calls and evaluating the
corresponding tool behavior against the user intent.

Compared with direct tracing by Codex and OpenClaw, \system improves
injection-point accuracy by 24.00\% and 25.00\%, respectively. \system also
improves path precision by 42.20\% and 42.98\%. These results indicate that
directly analyzing raw execution logs without a structured intermediate
representation makes it difficult to connect dispersed attack steps and
separate them from unrelated tool calls.

Compared with existing tracing methods, \system provides a better balance
between injection-point identification and attack-chain reconstruction.
AuthGraph achieves a higher detection rate but much lower injection-point
accuracy, while AgentArmor achieves higher path coverage but substantially
lower precision.

Overall, \system provides the strongest combined performance in injection-point
identification and precise attack-chain reconstruction.

\subsection{RQ3: Ablation Study}
\label{sec:rq3}

This experiment evaluates the contributions of the core tracing components. We randomly select 100 IPI attack chains (different from RQ2) within the historical-log
pool and evaluate the ablation setting. We construct the following three ablation settings.

\begin{itemize}[leftmargin=*]
  \item \textbf{Without \cspg Construction} replaces the \cspg with a temporal
  graph that directly connects adjacent tool calls.

  \item \textbf{Without User Intent Alignment} replaces
  the \uibs and alignment predicates with direct LLM judgments of whether each decision-grounded execution edge represents intent drift.

  \item \textbf{Without Event-guided Pruning} replaces event-guided pruning
  with full-history LLM analysis. The LLM analyzes the
  complete historical agent execution logs and determines the analysis, graph, and path scopes. This setting retains \cspg construction and user intent alignment.
\end{itemize}

\begin{table}[t]
  \caption{Ablation of the Core Tracing Components.}
  \label{tab:component-ablation}
  \centering
  \small
  \begin{tabular*}{\linewidth}
    {@{\extracolsep{\fill}}llcccc@{}}
    \toprule
    Configuration & Replacement & DetectionRate (\%) & InjectionAcc (\%)
      & Coverage (\%) & Precision (\%) \\
    \midrule
    W/o \cspg
      & Temporal graph
      & 78.00 & 74.00 & 58.54 & 80.97 \\
    W/o alignment
      & LLM judgment
      & 79.00 & 64.00 & 57.04 & 77.91 \\
    \system
      & None
      & \textbf{98.00} & \textbf{91.00}
      & \textbf{90.32} & \textbf{88.66} \\
    \bottomrule
  \end{tabular*}
\end{table}

Table~\ref{tab:component-ablation} reports the contributions of \cspg
construction and user-intent alignment to tracing effectiveness.
Table~\ref{tab:pruning-ablation} reports the effectiveness and efficiency of
tracing without event-guided pruning. The time and token columns report the additional cost of full-history LLM analysis relative to \system.

\begin{table}[t]
  \caption{Efficiency Impact of Ablating event-guided Pruning.}
  \label{tab:pruning-ablation}
  \centering
  \scriptsize
  \setlength{\tabcolsep}{3pt}
  \begin{tabular}{@{}llcccccc@{}}
    \toprule
    Configuration & Replacement & DetectionRate (\%) & InjectionAcc (\%)
      & Coverage (\%) & Precision (\%) & TimeInc. (\%) $\uparrow$ & TokenInc. (\%)$\uparrow$  \\
    \midrule
    W/o pruning
      & LLM analysis
      & 97.00 & 89.00 & 87.29 & 85.54 & 30.98 & 51.24 \\
    \bottomrule
  \end{tabular}
\end{table}

\noindent\textbf{Result Analysis.}
The results show that \cspg construction and user intent alignment
substantially improve tracing effectiveness, while event-guided pruning reduces
analysis cost with little loss of effectiveness.

Replacing the \cspg with a temporal graph reduces injection-point accuracy by
17.00\% and path coverage by 31.78\%. A temporal graph connects tool calls
according to execution order. Because tool calls driven by different task
intents may be interleaved, temporally adjacent calls may be unrelated. The
temporal graph therefore can not connect attack steps and introduces unrelated
connections. This result demonstrates that the \cspg is an effective
intermediate representation for IPI tracing.

Replacing user intent alignment with direct LLM judgments reduces
injection-point accuracy from 91.00\% to 64.00\%. Direct LLM judgments evaluate
agent behavior against natural-language user requests. However, user requests
are often vague and provide insufficient criteria for reliable intent
alignment. This result demonstrates the importance of structured authorization
criteria for identifying intent drift.

Full-history LLM analysis achieves tracing effectiveness close to that of the
full \system because \cspg construction and user intent alignment are
retained. However, full-history analysis increases average analysis time by
30.98\% and token consumption by 51.24\%. This result demonstrates that
event-guided pruning reduces analysis cost while preserving tracing
effectiveness.

\subsection{RQ4: User Intent Alignment on Benign Tasks}
\label{sec:rq4}

This experiment evaluates whether user intent alignment incorrectly flags
legitimate agent behavior. We select 100 additional injection-free agent
execution logs from AgentDyn. For each agent execution log, we construct the
\cspg and apply user intent alignment to every evaluable grounded \cspr. We do
not perform attack-chain traceback because these logs contain no
IPI-induced anomalous event, injection point, or annotated attack chain. We
measure the proportion of legitimate decision-grounded execution edges that
are incorrectly labeled \textsc{Misaligned}.

\noindent\textbf{Result Analysis.}
User-intent alignment achieves an edge-level false-positive rate of 3.52\% on
legitimate agent behavior. The 100 agent execution logs contain 394 evaluable
execution edges. Manual review shows that 25 edges labeled
\textsc{Misaligned} represent genuine execution deviations rather than false
positives. After excluding these edges, 13 of the remaining 369 legitimate
edges are incorrectly labeled \textsc{Misaligned}, producing the reported
false-positive rate.
The effect of an isolated false-positive edge is further limited because
\system attributes an anomalous event to IPI only when an intent-drift path
connects an \textit{ExternalResource} root node to the sink node.

\section{Discussion}
\label{sec:discussion}

\subsection{Applicability and Extensions}

\para{Applicable Agent Execution Paradigms}
\system is best suited to agents that follow ReAct or a similar
perceive--plan--act--observe loop \cite{yao2023react}. Such agents record
reasoning before tool calls, providing the evidence needed to extract
contextual decision relations and recover decision dependencies among tool
calls. A black-box agent without reasoning records can be analyzed only
through tool records. The temporal-graph ablation shows that tracing
effectiveness decreases when reasoning evidence is unavailable.

\para{Extension to Persistent IPI}
The design of \system can be extended to persistent IPI across user sessions.
In such attacks, malicious content introduced in an earlier session is stored
in memory, configuration, or another persistent entity and influences agent
behavior in a later session. When the root node of an intent-drift path
represents a persistent entity, \system can search historical agent execution
logs for the tool call that created or modified the entity. \system can then construct a \cspg for the earlier session and connect the earlier-session graph to the current \cspg through the shared persistent entity.

We reproduced 45 cross-session IPI cases from the stored prompt-injection study
by Xie et al. \cite{xie2026stored}. \system achieved more than 85\%
injection-point accuracy in this exploratory experiment. However, the sample
size and covered scenarios are insufficient to support a general conclusion.
We therefore exclude this result from the formal evaluation and leave a broader evaluation of persistent entities and cross-session tracing to future work.

\subsection{Threats to Validity}
\label{sec:validity}

\para{Internal Validity}
Some comparison methods have not released their complete implementations. We reproduce these methods by following the published algorithms, prompts, and rules. We provide all methods with the same comparison dataset, anomalous events, annotations, and LLM backbone when applicable to reduce differences introduced by the experimental setup. Our conclusions do not rely solely on these comparisons. The end-to-end and ablation experiments independently evaluate the effectiveness of \system and the contributions of its core components.

\para{External Validity}
First, \system currently focuses on tracing IPI expressed through natural language
context. \system does not directly trace code-level injection, in which malicious code may modify files, change environment variables, or issue network requests without exposing these behaviors as separate agent tool calls.
However, existing system-level tracing techniques can recover such behavior
from system calls, file accesses, and network activity
\cite{hossain2017sleuth,hassan2019nodoze,han2020unicorn}. In practice, \system can extend its coverage to code-level injection by integrating these techniques and linking their execution records to the agent tool call that launched the code.
This integration would allow \system to reconstruct the attack chain to code-level behavior.

Second, reproducing the log environment of a long-running industrial agent deployment is difficult because such an environment may accumulate vast volumes of execution data over time. We approximate this setting using a historical-log pool constructed from five months of Codex execution logs. In this environment, background tool calls outnumber attack-chain tool calls by approximately 3,000:1. The experimental results demonstrate that \system remains effective under substantial historical-log noise. This robustness follows from the event-guided pruning design: \system uses the anomalous event to restrict analysis to relevant log segments and intent-drift paths instead of constructing and examining a global graph over the complete log history.

\section{Related Work}
\label{sec:related-work}

\para{Prompt Injection Attacks and Benchmarks}
Prior work shows that instructions in external resources can redirect LLM
applications \cite{greshake2023ipi,liu2023houyi,liu2023promptinjection}.
Existing benchmarks cover prompt injection
\cite{toyer2023tensortrust,yi2023bipia}, agent tool execution
\cite{zhan2024injecagent,debenedetti2024agentdojo,zhang2024asb}, dynamic and
multi-step attacks \cite{li2026agentdyn,zhan2026stepjack}, and harmful
multi-tool tasks \cite{andriushchenko2024agentharm}. These studies demonstrate
that a malicious instruction can induce a sequence of tool calls, motivating the necessity to reconstruct complete attack chains

\para{Agent Safety Evaluation and Defenses}
Agent safety research covers both behavioral evaluation and
defenses at the model, system, and runtime levels.
Agent evaluations study interactive, web, and computer-use tasks
\cite{wang2023agentsurvey,liu2023agentbench,ma2024agentboard,
zhou2023webarena,xie2024osworld}, while safety benchmarks evaluate risky tool
use and harmful task completion
\cite{ruan2023toolemu,yuan2024rjudge,zhang2024agentsafetybench,
andriushchenko2024agentharm}. Existing defenses improve instruction separation
at the model level
\cite{wallace2024hierarchy,chen2024struq,chen2024secalign,
hines2024spotlighting}, enforce system-level information-flow or privilege
policies \cite{debenedetti2025camel,costa2025fides,shi2025progent}, or inspect
tool behavior at runtime
\cite{mou2026toolsafe,sequeira2026agentsentry,wang2026authgraph,
she2026provenanceguard}. In contrast, \system performs post-incident tracing to
locate the injection source and reconstruct the attack chain.

\para{Execution Tracing and Reasoning Evidence}
Execution-tracing research uses evidence recorded during execution to explain
observed behavior and identify the sources of that behavior.
PROV models entities, activities, and provenance relations
\cite{moreau2013prov}, while SLEUTH, NoDoze, and UNICORN use execution
provenance for security analysis
\cite{hossain2017sleuth,hassan2019nodoze,han2020unicorn}. Agent-tracing methods
model program dependencies, runtime states, capabilities, or authorization
relations \cite{wang2025agentarmor,li2026agentbom,liu2025traceaegis,
wang2026authgraph}; other methods identify influential sources through
contextual influence, counterfactual tests, or contribution signals
\cite{weng2026argus,cai2026neurotaint,jia2025promptlocate,
wang2025attntrace}. Section~\ref{sec:motivation} discusses their limitations
for multi-step IPI tracing.
Chain-of-thought, ReAct, and structural representations expose reasoning or
execution structure
\cite{wei2022cot,yao2023react,iyer2026structural}. Because generated reasoning
may be unfaithful \cite{turpin2023unfaithful}, \system grounds contextual
decision relations extracted from reasoning records in executed tool calls.


\section{Conclusion}
\label{sec:conclusion}

We presented \system, an intent-aware tracing framework for IPI attacks in
agent execution logs. \system treats IPI as a form of intent drift in
which a malicious instruction causes the agent's execution to deviate from the
user intent. \system constructs an Intent-Driven Execution Graph that
connects tool calls by task intent and aligns the recovered task intents with the user intent. The alignment identifies the
tool-call sequence that represents the attacker intent, allowing \system to
reconstruct the attack chain and locate the injection source and injection
point.
Experiments on datasets of agent execution logs constructed from AgentDyn and
InjecAgent demonstrate accurate injection-point identification and precise
attack-chain reconstruction. The reconstructed evidence supports incident
analysis and response.

\bibliographystyle{ACM-Reference-Format}
\bibliography{main}


\end{document}